\documentclass[sigconf]{acmart}

\AtBeginDocument{%
  \providecommand\BibTeX{{%
    \normalfont B\kern-0.5em{\scshape i\kern-0.25em b}\kern-0.8em\TeX}}}

\copyrightyear{2025} 
\acmYear{2025} 
\setcopyright{rightsretained}
\acmConference[C\&C '25]{Creativity and Cognition}{June 23--25, 2025}{Virtual, United Kingdom}
\acmBooktitle{Creativity and Cognition (C\&C '25), June 23--25, 2025, Virtual, United Kingdom}\acmDOI{10.1145/3698061.3734399}
\acmISBN{979-8-4007-1289-0/2025/06}

\begin{document}

\title{Tracing the Invisible: Understanding Students' Judgment in AI-Supported Design Work}

\author{Suchismita Naik}
\email{naik33@purdue.edu}
\orcid{0009-0002-5667-4576}
\affiliation{
  \institution{Purdue University}
  \city{West Lafayette}
  \state{Indiana}
  \country{USA}
}
\author{Prakash Shukla}
\email{shukla37@purdue.edu}
\orcid{0009-0002-7416-1758}
\affiliation{
  \institution{Purdue University}
  \city{West Lafayette}
  \state{Indiana}
  \country{USA}
}
\author{Ike Obi}
\email{obii@purdue.edu}
\orcid{0000-0002-3910-8890}
\affiliation{
  \institution{Purdue University}
  \city{West Lafayette}
  \state{Indiana}
  \country{USA}
}
\author{Jessica Backus}
\email{backus@purdue.edu}
\orcid{0009-0008-0863-8080}
\affiliation{
  \institution{Purdue University}
  \city{West Lafayette}
  \state{Indiana}
  \country{USA}
}
\author{Nancy Rasche}
\email{nrasche@purdue.edu}
\orcid{0000-0002-8042-9258}
\affiliation{
  \institution{Purdue University}
  \city{West Lafayette}
  \state{Indiana}
  \country{USA}
}
\author{Paul Parsons}
\email{parsonsp@purdue.edu}
\orcid{0000-0002-4179-9686}
\affiliation{
  \institution{Purdue University}
  \city{West Lafayette}
  \state{Indiana}
  \country{USA}
}

\renewcommand{\shortauthors}{Naik et al.}

\begin{abstract}
As generative AI tools become integrated into design workflows, students increasingly engage with these tools not just as aids, but as collaborators. This study analyzes reflections from 33 student teams in an HCI design course to examine the kinds of judgments students make when using AI tools. We found both established forms of design judgment (e.g., instrumental, appreciative, quality) and emergent types: agency-distribution judgment and reliability judgment. These new forms capture how students negotiate creative responsibility with AI and assess the trustworthiness of its outputs. Our findings suggest that generative AI introduces new layers of complexity into design reasoning, prompting students to reflect not only on what AI produces, but also on how and when to rely on it. By foregrounding these judgments, we offer a conceptual lens for understanding how students engage in co-creative sensemaking with AI in design contexts. 
\end{abstract}




\maketitle

\section{Introduction} 


The recent rise of AI tools in educational settings has generated widespread interest across disciplines. Design education too has started embracing these technologies, with students increasingly incorporating AI tools into their design projects \cite{lively_integrating_2023}. In response, researchers have begun exploring students' attitudes toward AI to better understand how these tools are being adopted in academic work \cite{zheng_charting_2024}. 

However, beyond capturing student attitudes, it is crucial to examine the judgment-making that occurs as students interact with AI tools throughout their design processes. The concept of judgment has been a long-standing topic of concern in the design literature, including interaction design, industrial design, architecture, and instructional design disciplines, where it is used to describe how designers navigate ambiguity and complexity \cite{gray_judgment_2015, parsons_design_2020, nelson_design_2003}. With the growing integration of AI tools in design workflows, recent work has begun to explore the potential of these tools to support or transform the design process itself \cite{lee_when_2025}, while also raising concerns about de-skilling and cognitive offloading in professional practice \cite{shukla_-skilling_2025}.

In this evolving context, it is no longer sufficient to examine only how students perceive AI. Instead, we must also attend to the judgments they make while engaging with these tools. Judgments about usefulness, appropriateness, quality, ethics, and trust, which are central to conceptualizing AI as a creative partner and rethinking how students use AI for generating ideas, wireframes, storyboards, documentation, and other design artifacts. Understanding these forms of judgment is crucial for uncovering the cognitive and metacognitive processes students engage in when collaborating with AI \cite{kelley_cognitive_2008, pintrich_students_1992, hutson_human-ai_2023}. Such insights are essential for designing instructional strategies that help in fostering critical reflection, responsible AI use, and creative agency in learners \cite{kim_are_2023, hutson_human-ai_2023}.

With this in mind, our research aims to identify the ``judgments'' that characterize student–AI collaboration in HCI and design projects to inform  AI literacy instruction. Here, we define judgments as the moments where students might have made evaluative decisions about different aspects of AI-generated design work. These judgments shed light on how students navigated the affordances and limitations of AI tools, not simply as users of a tool but as critical participants in a co-creative process. To guide this investigation, we pose the following research question: What forms of judgment do students engage in as they interact with AI tools during the creative process?


\section{Background}
\subsection{Design Judgment}
In design education, judgment is foundational \cite{nelson_design_2012}. Unlike procedural decision-making, which relies on established rules or fixed criteria, design judgment involves situated, interpretive, and value-laden reasoning \cite{nelson_design_2003, zhu_case_2020}. This allows designers to make sense of complex, uncertain contexts, weigh competing goals, and take action in the face of ambiguity \cite{demiral-uzan_instructional_2024}. Nelson and Stolterman define design judgment as \textit{``the ability to gain subconscious insights that have been abstracted from experiences and reflections, informed by situations that are complex, indeterminate, indefinable, and paradoxical''} \cite{nelson_design_2012}. They proposed a typology of design judgments (see Table~\ref{tab:judgment_types}). Leveraging this, researchers such as Gray et al. \cite{gray_judgment_2015} and Parsons et al. \cite{parsons_design_2020} have noted how design practitioners engage in complex, layered forms of judgment-making within real-world design practice. Within educational settings, attending to students’ judgment practices offers a lens into how design knowledge is being constructed in real time—how students are learning to think like designers, not just follow procedures \cite{demiral-uzan_instructional_2015}.

\begin{table}
  \centering
  \caption{Nelson \& Stolterman's typology of Judgments}
  \label{tab:judgment_types}
  \begin{tabular}{ll}
    \toprule
    Judgment&Definition\\
    \midrule
    Default & Instinctive, automatic judgment \\
    Offhand & Conscious recall of past judgments \\
    Appreciative & Prioritizing certain aspects over others \\
    Appearance & Evaluating overall aesthetic quality \\
    Quality & Assessing alignment with norms \\
    Instrumental & Selecting tools and methods for design goals \\
    Navigational & Strategizing the approach to problem-solving \\
    Compositional & Integrating elements for holistic coherence \\
    Connective & Linking design components in context \\
    Core & Reflecting personal values in decisions \\
  \bottomrule
\end{tabular}
\end{table}



\subsection{Student-AI Collaboration}
As generative AI tools become increasingly embedded in creative workflows, they introduce new kinds of complexity—blurring authorship, shifting agency, and producing outputs that require nuanced evaluation \cite{yu_ai_2025, wen_ai_2024}. Understanding the judgments students make in these contexts is crucial: it reveals how they are learning to reason, reflect, and compose meaning within human–AI collaborations \cite{kothari_enhancing_2024}. Rather than treating AI use as a technical skill, we argue that educators must attend to students’ judgment-making practices as a site of design learning itself—and as a necessary foundation for cultivating critical, ethical, and autonomous designers in an AI-mediated world \cite{kumar_guiding_2024}.

\section{Methodology}

This study was conducted in an undergraduate UX Design course at a large Midwestern U.S. university, involving 175 students across four sections. Students worked in 35 teams on a design project with optional AI tool use, supported by consistent instructor guidance and clear expectations for citation and reflection. Of the students who submitted usable reflections, 52\% were freshmen, 20\% sophomores, 18\% juniors, and 10\% seniors. Participants represented a range of majors, including User Experience Design (29\%), Game Design \& Development (25\%), Animation \& Video Effects (12\%), and others. The students were given a design project prompt for designing a \textit{fitness tracking experience} which was followed by explicit instruction as:

\begin{quote}
    \textit{USE OF GENERATIVE AI. In this project, you are encouraged to leverage Generative AI tools, including LLMs, in your design process. \textbf{Be sure to justify and provide a rationale for their use; give credit to the AI tools with a citation and explanation of how you used the tool}. Include a group reflection (300-500 words) in your documentation discussing your experiences of using these tools in your project, highlighting both their advantages and challenges. Your use of Generative AI and the contents of your group reflection will not negatively impact your Project 3 scores, and for this project, you'll be exempt from the AI usage policy.}
\end{quote}

Data consisted of group and individual reflections on AI usage, documented in the project submissions of 33 out of 35 teams. The two teams lacking reflection sections were excluded from the analysis. A thematic analysis informed a three-part codebook (content type, design phase, design judgment types). Five researchers, including the instructors, used this framework in a structured content analysis to identify patterns in design judgment around AI practices, using collaborative tools and regular discussions to ensure consistency and reliability.

\section{Findings}
This section presents our initial findings on how undergraduate design students exercised judgment when interacting with generative AI tools in the context of a course-based design project. Drawing from 33 group reflections, we identified six key categories of judgments - moments where students made evaluative decisions about the use, appropriateness, quality, or implications of AI-generated content. The categories we identified are (highlighted in table ~\ref{tab:summary_judgment}): judgments about AI's (1) reliability, (2) task appropriateness, (3) evaluations of output quality, (4) iterative refinement and calibration, (5) ethical and epistemic considerations, and (6) role interpretation as tool or collaborator.

\begin{table*}[t]
  \caption{Summary of Judgment Categories Identified in Student Reflections}
  \label{tab:summary_judgment}
  \begin{tabular}{
    p{0.30\linewidth} 
    p{0.20\linewidth} 
    p{0.20\linewidth}
    p{0.20\linewidth}
  }
    \toprule
    \textbf{AI-Supported Design Judgment Category} & \textbf{Judgment Description} & \textbf{Representative Examples} & \textbf{Overlap with Nelson \& Stolterman's Judgment Typology} \\
    \midrule

    \textbf{Reliability Judgments} & 
    Assessing trustworthiness, factual accuracy, or consistency & ``Not always trustworthy,'' ``needed human review'' & -- \\

    \midrule
    \textbf{Appropriateness Judgments} & 
    Evaluating task fit or usefulness based on context & 
    ``Too complex for AI,'' ``only used for low-stakes components'' & Appreciative Judgment \\

    \midrule
    \textbf{Output Quality Judgments} & 
    Judging creativity, usability, or clarity of generated content & ``Fluff,'' ``uninspired but sparked ideas,'' ``wrong tone/style'' & Quality \& Appearance Judgment \\

    \midrule
    \textbf{Calibration \& Refinement Judgments} & 
    Iterative prompt revision and output tweaking & 
    ``Revised inputs,'' ``adjusted to make it work better'' & Instrumental \& Navigational Judgment\\

    \midrule
    \textbf{Ethical/Epistemic Judgments} & 
    Reflecting on integrity, originality, and ethical use & 
    ``Didn't feel right,'' ``borrowed from other work,'' ``opted out'' & -- \\

    \midrule
    \textbf{Agency Distribution Judgments} & 
    Conceptualizing AI as tool, partner, or authority figure & 
    ``Helpful assistant,'' ``settled team debates,'' & -- \\

    \bottomrule
  \end{tabular}
\end{table*}

\subsection{Judgments of Reliability and Trustworthiness}
A frequent form of evaluative reasoning centered on the perceived reliability and credibility of AI outputs. Students demonstrated an awareness of the inherent limitations of generative AI, particularly with regard to factual accuracy, recency of knowledge, and susceptibility to hallucinations or bias.

For example, Team 4 reflected, \textit{``AI isn't completely reliable… we needed to double-check and revise the work,''} while Team 6 noted, \textit{``ChatGPT is still very prone to errors... this keeps the human incorporation intact.''} Similarly, Team 16 observed that \textit{``the ideas it gave us weren't the most useful... they just backed up research we already had,''} and Team 23 elaborated, \textit{``It was empowering and humbling to witness the capabilities of machine learning algorithms... but we encountered challenges when relying on outdated or incomplete sources.''}

These reflections illustrate that students were not treating AI-generated content as authoritative or infallible. Instead, they applied human judgment to assess the credibility of the output, identifying moments where further verification, revision, or supplementation was required. This mode of engagement reflects an emergent critical literacy around AI tools, in which students recognize the importance of human oversight in maintaining accuracy and relevance.

\subsection{Judgments of Appropriateness for Specific Tasks}
In addition to assessing reliability, students made decisions about whether AI was suitable for the task at hand. These judgments reflect a nuanced understanding of the affordances and constraints of AI tools across different design phases.

For instance, Team 1 explicitly rejected the use of AI-generated imagery due to a mismatch with task goals, while Team 4 noted, \textit{``It is very difficult to create optimal AI artworks... asking AI to create something that clearly presents our ideas is nearly impossible.''} In contrast, Team 5 reported using AI for logo generation, citing time constraints and the complexity of logo design as justification: \textit{``Designing a logo is complicated and we didn't have time.''} Similarly, Team 18 limited AI use to low-stakes components, stating, \textit{``We wouldn't want to use it for something more significant than [naming the app].''}

These reflections point to contextual judgments, where students evaluated whether AI added value based on the specificity, creative complexity, or stakes of the design task. In doing so, students exhibited task sensitivity-recognizing, for example, that AI may support rapid iteration in early-stage ideation but may be ill-suited for highly expressive or nuanced visual work.

\subsection{Judgments of Output Quality and Usability}
A third category of judgments focused on the quality, creativity, or stylistic appearances of AI outputs. These evaluations extended beyond correctness to include more subjective or aesthetic criteria, which are central to design work. Team 2 described AI-generated ideas as \textit{``great for expanding early ideation but [they] need vetting for biases.''} Team 10 expressed frustration with image-generation tools: \textit{``Image generation was difficult... colors were off, style wasn't right.''} Team 24 reflected on the mixed utility of AI-written summaries and protocols, noting that \textit{``AI often added fluff,''} while Team 12 found that \textit{``names generated were uninspired, but sparked inspiration.''}

Students, in these cases, demonstrated a discerning approach to output evaluation, identifying both the benefits and the limitations of AI contributions. Importantly, some teams noted that even low-quality or imperfect outputs could serve as creative prompts, indicating that students did not always measure success by output fidelity alone, but also by the generative value of AI in spurring human ideation.

\subsection{Judgments Through Iterative Refinement and Calibration}
Another significant judgment pattern involved iterative engagement with AI tools, where students actively adjusted prompts, refined inputs, or reworked outputs in response to AI-generated content. This calibration process reflects a deeper understanding of AI as a system that can be influenced through deliberate human interaction.

Team 1 reported that they \textit{``trimmed article content to make ChatGPT work better,''} suggesting an awareness of input optimization. Team 3 described a multi-step workflow: \textit{``Used AI to draft an outline, then used human input to elaborate,''} while Team 14 engaged in prompt revision to improve output fit: \textit{``Revised prompts for better fit after AI gave unhelpful suggestions.''}

These interactions suggest that students were learning to shape the behavior of AI through prompt engineering and iterative refinement. Rather than treating AI outputs as static or final, students engaged in a dynamic process of co-construction, reflecting a growing proficiency in managing the affordances of generative tools.

\subsection{Ethical and Epistemic Judgments}
Beyond functional concerns, several teams raised questions about the ethical, creative, or philosophical implications of using AI in the design process. These reflections foregrounded students' internal value systems and their evolving sense of professional responsibility.

Team 18 expressed concern about the originality of AI-generated content: \textit{``We had reservations using AI, as it borrows from other work. We limited use to naming.''} Teams 27 and 33 independently chose not to use AI at all, citing ethical or value-based concerns such as bias, privacy, or the risk of displacing human creativity. Meanwhile, Team 24 documented a change in attitude, moving from skepticism to conditional acceptance: \textit{``Shifted from viewing AI as cheating to a legitimate support tool.''}

These judgments reveal that students were not simply weighing effectiveness but were also interrogating the legitimacy and appropriateness of AI's presence in the design process. Such epistemic and ethical reflections highlight the importance of integrating critical discourse about AI's societal and disciplinary implications into design education curricula.

\subsection{Judgments About Agency Distribution}
Students held varied views on AI's role in the design process, assigning it agency as a tool, collaborator, or authority. These mental models influenced how students interacted with the technology, the degree of autonomy they granted it, and the kinds of tasks they felt were appropriate for AI to perform.

Team 6 described AI as a \textit{``helpful tool but kept human oversight for credibility,''} indicating a tool-based mental model. In contrast, Team 25 treated AI as a quasi-authoritative third party, using it to mediate group decision-making: \textit{``AI is a `third-party' to settle group decisions.''} Team 17 creatively deployed AI to generate content with specific tones: \textit{``Used AI for crafting tone-specific content like humorous prompts and `sassy' messages.''}

These illustrate that students' perceptions of AI ranged from that of an instrumental assistant to a co-creative partner or even an adjudicator in team dynamics. Such interpretations can profoundly influence how students evaluate outputs and assign responsibility, underscoring the need for explicit discussion about AI's role in collaborative and creative work.

\section{Discussion}
Our analysis indicates that students engaged in a diverse range of design judgments while working with generative AI. Many of these align with Nelson and Stolterman’s typology of design judgment \cite{nelson_design_2012}, as well as with forms of judgment observed among UX practitioners in real-world settings \cite{shukla_coping_2025}, including appreciative, instrumental, appearance, quality, and navigational judgments. For example, task-appropriateness judgments reflected instrumental and navigational reasoning, while output evaluation drew on aesthetic and appearance-based criteria.

However, we also identified two new forms of judgments that extend beyond existing frameworks: \textit{agency-distribution judgment} and \textit{reliability judgment.}

\textbf{Agency-distribution judgment} captures how students conceptualize and negotiate the division of creative responsibility between themselves and AI. This aligns with Guo et al.'s work, which explores how designers perform, shift, or suppress agency when co-creating with AI, depending on the stage of ideation and the cognitive load introduced by AI tools \cite{guo_rethinking_2023}. Students in our study often assigned AI the role of a prompt generator, a consensus builder, or an ``extra team member,'' but they also expressed tensions about ceding too much authorship. These findings echo recent calls in HCI to examine how AI shapes not only outcomes but also who gets to design and how responsibility is distributed \cite{stoimenova_exploring_2020}.

\textbf{Reliability judgment} describes students' evaluation of the trustworthiness and relevance of AI outputs—beyond stylistic quality—particularly regarding factual accuracy and appropriateness.

These new judgments show how generative AI tools add layers of metacognitive work to design process---prompting reflection not just on the product, but on the tool’s role and influence. This broadens the landscape of design judgment by foregrounding how cognition and responsibility are shared across humans and AI in co-creative contexts. For educators, recognizing these judgments has practical value. Instead of emphasizing tool proficiency alone, instruction should scaffold reflective practice: helping students decide when to use AI, evaluate its outputs, and maintain creative autonomy.


\section{Limitations and Future Work}
Due to the inherent methodological constraints of this study, particularly the collection of student reflections only toward the project's conclusion, we were unable to thoroughly capture or investigate certain types of judgments. Specifically, \textit{default judgments, core judgments,} and \textit{compositional judgments} were not fully addressed within the scope of our analysis. Also, this study focuses on a single UX design course within one discipline at a large Midwestern university in the US, which may limit the generalization of the findings. 

To address this limitation, we plan to extend the research across additional UX design courses over time to build upon these preliminary insights. Future phases will adopt more rigorous methods, such as think-aloud protocols \cite{atman_verbal_1998}, design session recordings \cite{christensen_studying_2014}, in-depth interviews, and focus groups \cite{crandall_working_2006} with interested participants. In future work, we aim to design activities that will allow for a more thorough exploration of the judgment types that were not captured in the present study. These efforts aim to deepen our understanding of students' cognitive processes in conceptualizing AI within collaborative creative practices, and how AI influences their design preferences and nuanced judgment decisions.

In addition, we aim to develop pedagogical frameworks that support critical AI literacy within design education. These frameworks should help students not only engage with generative AI tools but also reflect critically on the ethical, epistemic, and creative implications of their use. By making judgment-making processes explicit, educators can better scaffold students' ability to evaluate AI outputs, calibrate tool use, and navigate complex design choices with greater autonomy and awareness.


\section{Contribution}
This research advances understanding of how novice designers critically engage with generative AI tools during the creative process. By identifying six categories of ``judgments,'' we offer a vocabulary and conceptual framework for analyzing how students make evaluative decisions in human-AI co-creativity. These findings move beyond tool use to surface the cognitive and ethical dimensions of AI-supported design work. Aligned with the conference's theme, \textit{Creativity for Change}, with this study we invite new pedagogical strategies for responsibly integrating AI in design education, equipping emerging designers with the critical AI literacy needed to navigate evolving generative AI technologies and contribute to socially conscious, reflective creative practice.

\begin{acks}
This work was supported by NSF award \#2146228.
\end{acks}

\bibliographystyle{ACM-Reference-Format}
\bibliography{references}

\end{document}